\documentclass{PoS}

\title{NNLO QCD analysis of the virtual photon structure functions}

\ShortTitle{NNLO QCD analysis of the virtual photon structure functions}

\author{Ken Sasaki\\
        Dept. of Physics, Faculty of Engineering, Yokohama National University\\        Yokohama 240-8501, Japan\\
        E-mail: \email{sasaki@phys.ynu.ac.jp}}

\author{Takahiro Ueda\\
        High Energy Accelerator Research Organization (KEK)\\
        1-1 Oho, Tsukuba, Ibaraki 305-0801, Japan\\
        E-mail: \email{uedat@post.kek.jp}}

\author{Yoshio Kitadono\\
        Dept. of Physics, Faculty of Science, Hiroshima University\\
        Higashi Hiroshima 739-8526, Japan\\
        E-mail: \email{kitadono@scphys.kyoto-u.ac.jp}}

\author{\speaker{Tsuneo Uematsu}\\
        Dept. of Physics, Graduate School of Science, Kyoto University\\
        Yoshida, Kyoto 606-8501, Japan\\
        E-mail: \email{uematsu@scphys.kyoto-u.ac.jp}}

\abstract{The next-to-next-to-leading order (NNLO) QCD analysis is 
performed for the virtual photon structure functions which can be 
measured in the double-tag events in two-photon processes in $e^+e^-$
collisions. We investigate the perturbative QCD evaluation of 
$F_2^\gamma(x,Q^2,P^2)$ to NNLO and $F_L^\gamma(x,Q^2,P^2)$ to NLO 
with and without taking into account the target mass effects, which 
are relevant for the large $x$ region. We also carry out 
the phenomenological analysis for the experimentally
accessible effective structure function $F_{\rm eff}^\gamma
=F_2^\gamma+(3/2)F_L^\gamma$. 
}

\FullConference{8th International Symposium on Radiative Corrections \\
		 October 1-5, 2007\\
		 Florence, Italy}

\begin{document}

\section{Introduction}
The perturbative QCD is now in a stage of improving its predictions more 
precise, in order to perform the accurate estimation of the strong 
interaction effects at LHC processes. What I would like to discuss in this 
talk is the next-to-next-to-leading order QCD analysis of the virtual photon 
structure functions. We consider the photon structure functions $F_2^\gamma$ 
and $F_L^\gamma$ which can be measured in the two-photon process in $e^+e^-$
collision
at high energies. In the future new kinematical regime is expected to be
available in the linear collider, ILC. Here we investigate
the double-tag events where both of the outgoing $e^+$ and $e^-$ are
detected, then one can study the virtual photon structure functions,
which describe the deep inelastic scattering off the virtual photon. 
In particular we consider the kinematical region in which one of the
photon (\lq probe\rq photon) is far off-shell while the other one
(\lq target\rq photon) is less off-shell but much bigger than the QCD
scale parameter $\Lambda$:
\begin{equation}
\Lambda^2 \ll P^2 \ll Q^2~, \label{Kinematical region}
\end{equation}
where $Q^2$ ($P^2$) is the mass squared of the
probe (target) photon.
% and $\Lambda$ is the QCD scale parameter.  
In this situation the structure functions are perturbatively calculable
not merely $Q^2$ dependence but also the shape and magnitude.  

In the framework based on the operator product expansion (OPE) 
supplemented by the renormalization (RG) group method, Witten 
\cite{Witten} obtained the leading order (LO) QCD contributions to 
$F_2^\gamma$ and $F_L^\gamma$ and, shortly after,  
the next-to-leading order (NLO) QCD corrections to $F_2^\gamma$ were
calculated by Bardeen and Buras \cite{BB}. 
The structure functions $F_2^\gamma(x,Q^2,P^2)$ 
and $F_L^\gamma(x,Q^2,P^2)$ for the case of a virtual photon target ($P^2\ne 0$)were studied in the  LO and in the
NLO by pQCD~\cite{UW}. In ref.\cite{USU}, we have studied 
the unpolarized virtual photon structure functions, 
$F_2^\gamma(x,Q^2,P^2)$ up to the NNLO and $F_L^\gamma(x,Q^2,P^2)$ 
up to the NLO,  in pQCD for the kinematical region (\ref{Kinematical
region}).  
In this talk we present the perturbative QCD evaluation of the above
structure functions with and without taking into account the target 
mass effects. We also discuss the experimentally accessible effective 
structure function $F_{\rm eff}^\gamma=F_2^\gamma+(3/2)F_L^\gamma$. 

\section{Virtual photon structure functions $F_2^\gamma(x,Q^2,P^2)$ 
and $F_L^\gamma(x,Q^2,P^2)$}

We analyze the virtual photon structure functions 
$F_2^{\gamma}(x,Q^2,P^2)$ and $F_L^{\gamma}(x,Q^2,P^2)$ using the
theoretical framework based on the OPE and the RG method. 
The absorptive part of the forward virtual photon scattering amplitude 
for $\gamma(q)+\gamma(p)\rightarrow \gamma(q)+\gamma(p)$ is 
related to the structure tensor:
\begin{eqnarray}
W_{\mu\nu}^\gamma(p,q)
=\frac{1}{2}\int d^4 x e^{iqx}\langle \gamma (p)|J_\mu(x) J_\nu(0)|\gamma (p)\rangle_{\rm spin\  av}.~,
\label{Wmunu}
\end{eqnarray}
which is expressed in terms of two independent structure functions 
\begin{eqnarray}
W_{\mu\nu}^\gamma=e_{\mu\nu}\left\{\frac{1}{x}
F_L^\gamma+\frac{p^2q^2}{(p\cdot q)^2}\frac{1}{x}F_2^\gamma
\right\}
+d_{\mu\nu}\frac{1}{x}F_2^\gamma~, 
\end{eqnarray}
where 
we have kept the target mass squared $p^2=-P^2$ and 
the two tensor structures are given by
\begin{eqnarray}
e_{\mu\nu}\equiv g_{\mu\nu}-\frac{q_\mu q_\nu}{q^2}, \quad
d_{\mu\nu}\equiv
-g_{\mu\nu}
+\frac{p_\mu q_\nu+p_\nu q_\mu}{p\cdot q}
- \frac{p_\mu p_\nu q^2}{(p\cdot q)^2}
\end{eqnarray}
and $x$ is the Bjorken variable defined 
by $x=Q^2/2p\cdot q$ with $q^2=-Q^2$. 

%%%%%%%%%%%%%%%%%%%%%%%%%%%%%%%%%%%
\begin{figure*}
\centering
  \begin{tabular}{c@{\hspace*{10mm}}c}
    \includegraphics[scale=0.3]{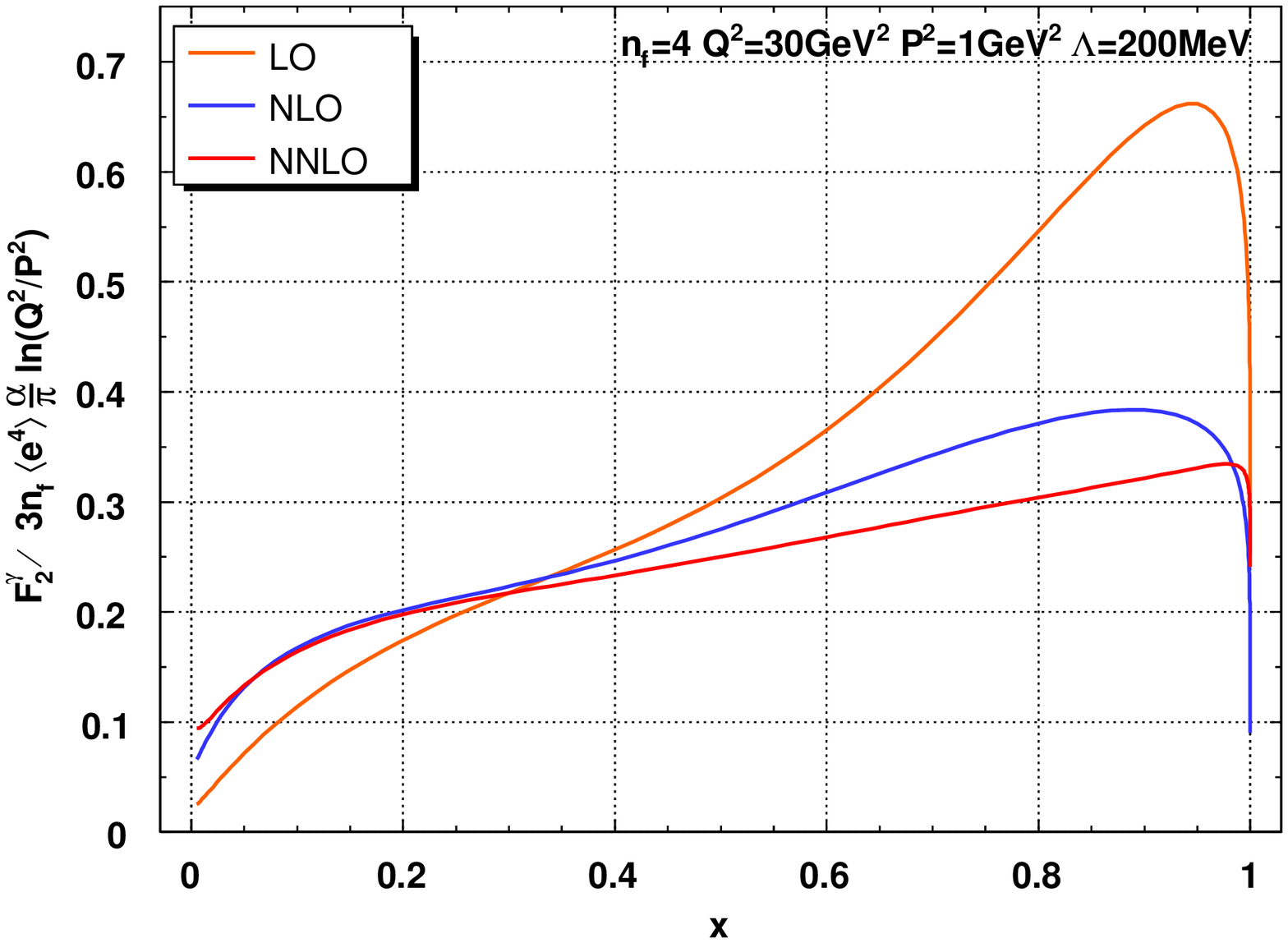}
    &
   \includegraphics[scale=0.3]{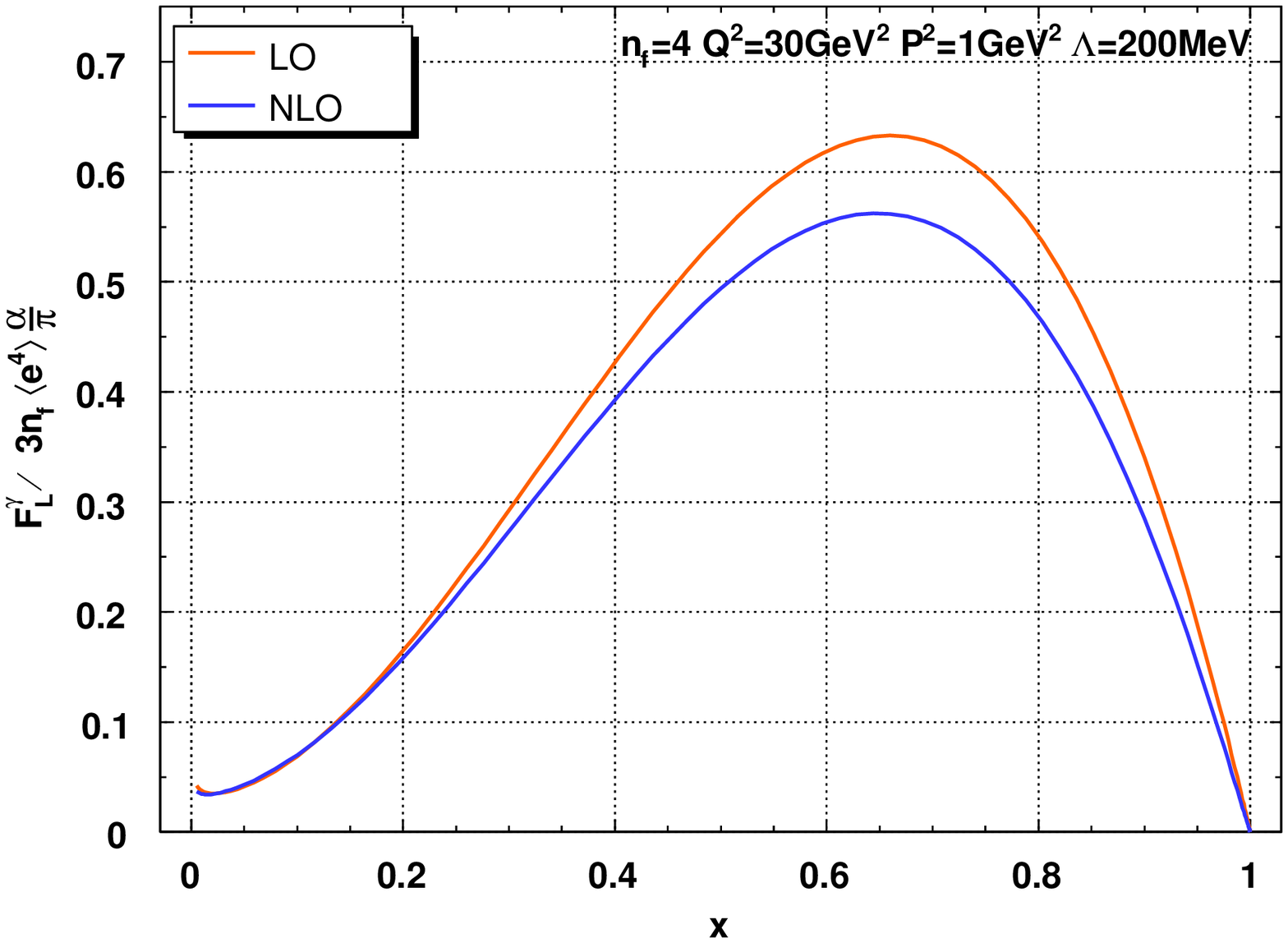}\\
(a) & (b)
  \end{tabular}
\caption{(a)$F_2^\gamma(x,Q^2,P^2)$ to NNLO 
and (b) $F_L^\gamma(x,Q^2,P^2)$ to NLO for $Q^2=30$GeV$^2$, $P^2=1$GeV$^2$ 
with $\Lambda=0.2$GeV.
}
\label{fig1}
\end{figure*}
%%%%%%%%%%%%%%%%%%%%%%%%%%%%%%%%%%%
The $n$-th moment of $F_2^\gamma(x,Q^2,P^2)$ up to NNLO 
without the target mass effects is given as
\begin{eqnarray}
&&M_{2,n}^\gamma(Q^2,P^2)\equiv \int_0^1 dx x^{n-2}F_2^\gamma(x,Q^2,P^2) 
\nonumber\\
&&=\frac{\alpha}{4\pi}\frac{1}{2\beta_0}
\Biggl\{\frac{4\pi}{\alpha_s(Q^2)}\sum_{i}{\cal L}^n_i
\left[1-\left(\frac{\alpha_s(Q^2)}{\alpha_s(P^2)}\right)^{d_i^n+1}
\right]+\sum_{i}{\cal
A}_i^n\left[1-\left(\frac{\alpha_s(Q^2)}{\alpha_s(P^2)}\right)^{d_i^n}\right]
\nonumber\\
&&\hspace{1cm}+\sum_{i}{\cal
B}_i^n\left[1-\left(\frac{\alpha_s(Q^2)}{\alpha_s(P^2)}\right)^{d_i^n+1}\right]
+{\cal C}^n
+\frac{\alpha_s(Q^2)}{4\pi}\Biggl(\sum_{i}{\cal
D}_i^n\left[1-\left(\frac{\alpha_s(Q^2)}{\alpha_s(P^2)}\right)^{d_i^n-1}\right]
\nonumber\\
&&\hspace{1cm}+\sum_{i}{\cal
E}_i^n\left[1-\left(\frac{\alpha_s(Q^2)}{\alpha_s(P^2)}\right)^{d_i^n}\right]
+\sum_{i}{\cal
F}_i^n\left[1-\left(\frac{\alpha_s(Q^2)}{\alpha_s(P^2)}\right)^{d_i^n+1}\right]
+{\cal G}^n \Biggr) 
+{\cal O}(\alpha_s^2)
 \Biggr\}~,
\label{master1}
\end{eqnarray}
where the index $i$ runs over $+, -, NS$ and $d_i^n=\lambda_i^n/2\beta_0$ and 
$\lambda_i^n$ denotes the eigenvalues of 1-loop anomalous dimension matrices. 
The terms with ${\cal L}^n_i$ are the LO ($\alpha/\alpha_s$) contributions 
\cite{Witten,UW}. The NLO ($\alpha$) corrections are the terms with 
${\cal A}^n_i$, ${\cal B}^n_i$ and ${\cal C}^n$ \cite{BB,UW}. 
The coefficients  ${\cal D}^n_i$, ${\cal E}^n_i$,  ${\cal F}^n_i$ and 
${\cal G}^n$ give the NNLO ($\alpha\alpha_s$) corrections and they are new. 
The explicit expressions of ${\cal D}^n_i$, ${\cal E}^n_i$,  ${\cal F}^n_i$ 
and  ${\cal G}^n$ are given in Eqs.(2.34)-(2.37) of Ref.\cite{USU}.
For the 3-loop anomalous dimensions, we could use the recently calculated results of the three-loop anomalous dimensions for the quark and gluon 
operators and of the three-loop photon-quark and photon-gluon splitting 
functions \cite{MVV}. 
For the longitudinal structure function $F_L^\gamma(x,Q^2,P^2)$ we can 
similarly derive the $n$-th moment up to NLO. The NLO terms are 
($\alpha\alpha_s$) corrections which are new results, given in Eqs.(6.3)-(6.8) 
of Ref.\cite{USU}.

The LO, NLO and NNLO QCD results, as well as the box contribution,  
for the case of $F_2^\gamma(x,Q^2,P^2)$ $\left(F_L^\gamma(x,Q^2,P^2)\right)$ 
at $Q^2=30$ GeV$^2$ and $P^2=1$ GeV$^2$ with $n_f=4$, are shown in Fig.1(a) (Fig.1.(b)). 
For $F_2^\gamma$ we observe that there exist notable NNLO QCD corrections at 
larger $x$. The corrections are negative and the NNLO curve comes below the 
NLO one for $0.3\!\lesssim\!x\!<\!1$. 
We see from Fig.1(b) that 
the NLO QCD corrections for $F_L^\gamma$ are negative and the NLO curve 
comes below the LO one in the region  $0.2\!\lesssim\! x<1$.

\section{Target mass effects}
If the target is a real photon ($P^2=0$), there is no need to consider
target mass corrections. But when the target becomes off-shell, and for
relatively low $Q^2$,
we need to take into account target mass effects (TME).
TME is important also 
by another reason.  For the virtual photon target, the maximal value of
the Bjorken variable $x$  is not 1 but $x_{\rm max}={1}/(1+\frac{P^2}{Q^2})$,
due to the constraint $(p+q)^2 \ge 0$, which is in contrast to  
the nucleon case where $ x_{\rm max}= 1$.
The structure functions should 
vanish at $x=x_{\rm max}$. 
However the numerical analysis in Fig.1 shows that the predicted graphs do not
vanish but remains finite at $x=x_{\rm max}$. 
This flaw is due to the fact that TME have not been taken into account
 in the analysis.  
TME can be treated by considering the Nachtmann moments~\cite{NACHT}
to extract the definite spin contribution by expanding the amplitude in 
Gegenbauer polynomials. 
%%%%%%%%%%%%%%%%%%%%%%%%%%%%%%%%%%%
\begin{figure*}
\centering
  \begin{tabular}{c@{\hspace*{20mm}}c}
    \includegraphics[scale=0.45]{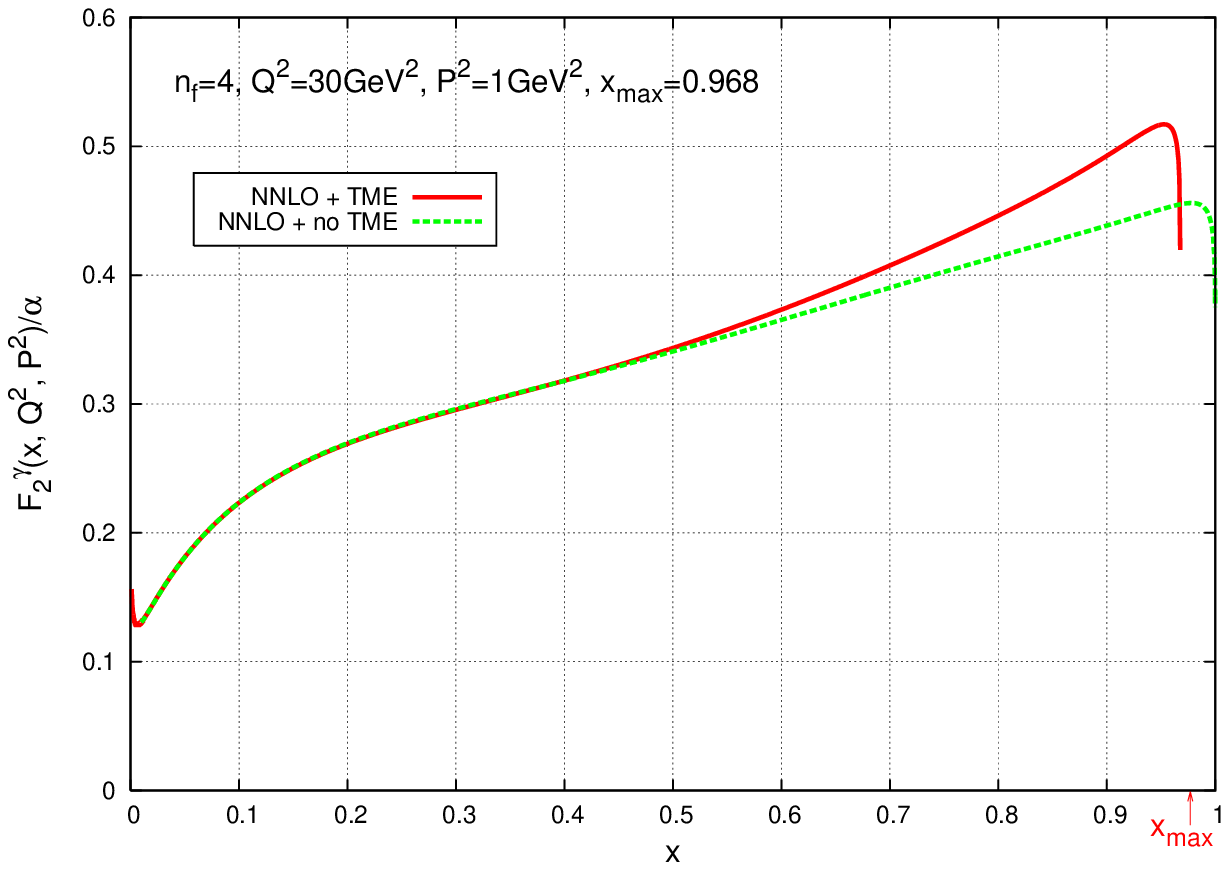}
    &
   \includegraphics[scale=0.45]{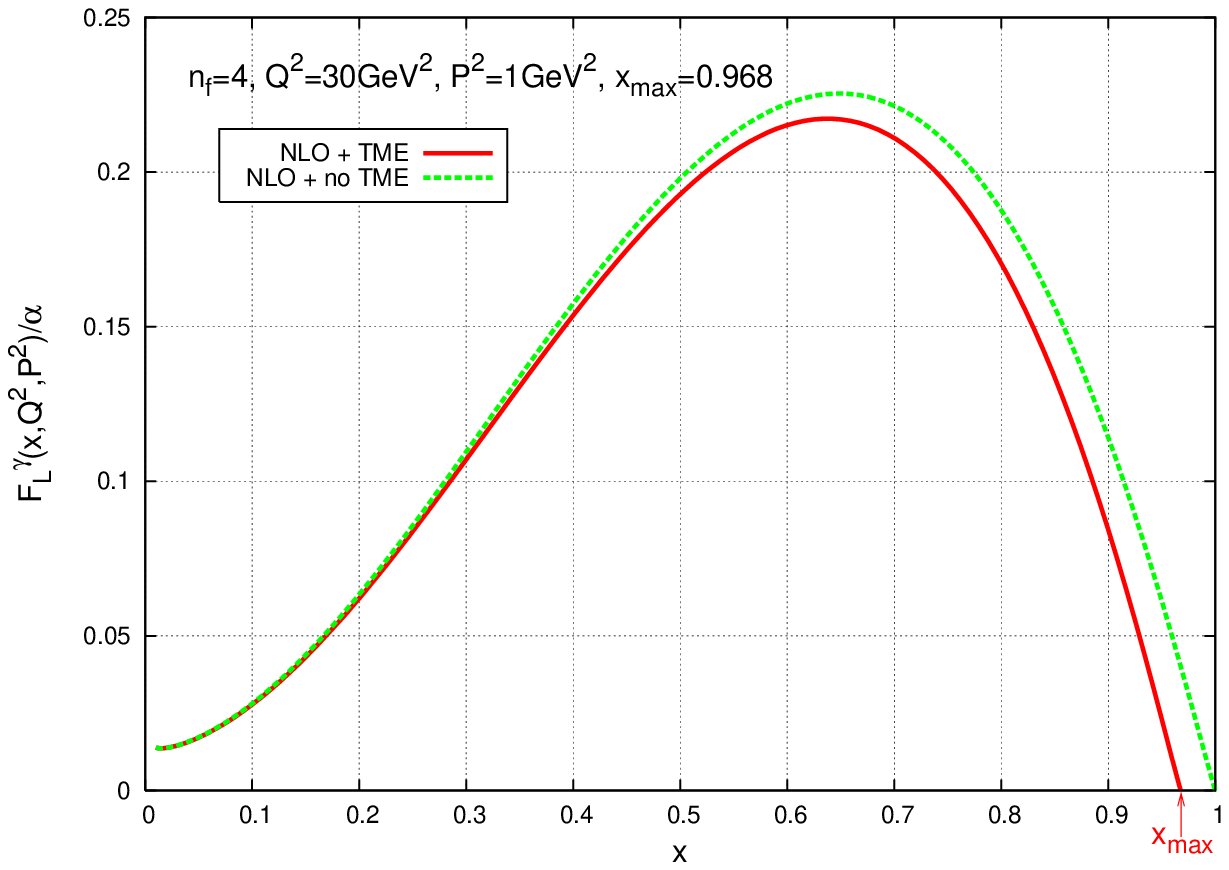}\\
(a) & (b)
  \end{tabular}
\caption{(a) $F_2^\gamma(x,Q^2,P^2)$ with TME to NNLO 
and (b) $F_L^\gamma(x,Q^2,P^2)$ with TME to NLO 
}
\label{fig2}
\end{figure*}
%%%%%%%%%%%%%%%%%%%%%%%%%%%%%%%%%%%
The Nachtmann moments for the definite spin-$n$ contributions, $M_{2,n}^\gamma$
and $M_{L,n}^\gamma$ are given by \cite{KSUU}
\begin{eqnarray}
&&\mu_{2,n}^\gamma(Q^2,P^2)\equiv
\int_0^{x_{\rm max}}dx\frac{1}{x^3}\xi^{n+1}
\left[
\frac{3+3(n+1)r+n(n+2)r^2}{(n+2)(n+3)}\right]
F_2^\gamma(x,Q^2,P^2)=M_{2,n}^\gamma\label{NachtmannM_2}\\
&&\mu_{L,n}^\gamma(Q^2,P^2)\equiv
\int_0^{x_{\rm max}} dx 
\frac{1}{x^3}\xi^{n+1}\left[F_L^\gamma(x,Q^2,P^2)\right.\nonumber\\
&&\hspace{4cm}\left.+\frac{4P^2 x^2}{Q^2}
\frac{(n+3)-(n+1)\xi^2 P^2/Q^2}{(n+2)(n+3)}F_2^\gamma(x,Q^2,P^2)\right]
=M_{L,n}^\gamma.
\label{NachtmannM_L}
\end{eqnarray}
where $\xi$ is the so-called $\xi$-scaling variable:
$\xi={2x}/(1+r)$, with $r=\sqrt{1-{4P^2x^2}/{Q^2}}$.
Inverting the Nachtmann moments to get the structure function
$F_{2(L)}^\gamma(x,Q^2,P^2)$ as a function of $x$, we have
\begin{eqnarray}
&&F_2^\gamma(x,Q^2,P^2)=\frac{x^2}{r^3}F(\xi)-6\kappa\frac{x^3}{r^4}H(\xi)
+12\kappa^2\frac{x^4}{r^5}G(\xi)\\ 
&&F_L^\gamma(x,Q^2,P^2)
=\frac{x^2}{r}F_L(\xi)-4\kappa
\frac{x^3}{r^2}H(\xi)+8\kappa^2\frac{x^4}{r^3}G(\xi)
\end{eqnarray}
where $\kappa=P^2/Q^2$.
The four functions $F(\xi)$, $H(\xi)$, $G(\xi)$, and $F_L(\xi)$ are given
by the inverse Mellin transforms of the moments $M_{2,n}^\gamma$ and
$M_{L,n}^\gamma$ divided by certain powers of $n$ as 
given in Ref.\cite{KSUU}.

We have plotted the $F_2^\gamma$ ($F_L^\gamma$) with and without
TME in Fig.2(a) $\left({\rm Fig2.(b)}\right)$ for 
$Q^2=30$GeV$^2$, $P^2=1$GeV$^2$.
We observe that TME become sizable at larger
$x$ region. While TME enhances $F_2^\gamma$ at larger $x$, it reduces 
$F_L^\gamma$.  In fact, $F_2^\gamma$ becomes maximum at $x$ very close 
to the maximal value of $x$, $x_{\rm max}$ (1) for the case with (without) TME.
In the case of $F_L^\gamma$ the maximum is attained at middle $x$.

\section{Concluding remarks}

Finally let us compare our theoretical prediction for the virtual photon 
structure functions with the existing experimental data.
In Fig.3(a) and Fig.3(b), we have plotted the experimental data from PLUTO 
Collaboration \cite{PLUTO} and also those from L3 Collaboration \cite{L3} 
on the so-called \lq\lq effective photon structure function\rq\rq\ defined 
as $F_{\rm eff}^\gamma=F_2^\gamma+\frac{3}{2}F_L^\gamma$, together
with the theoretical predictions, NNLO QCD with and without TME
and also Box diagram calculations. 
Although the experimental error bars are rather large, the data
are considered to be roughly consistent with the theoretical expectations,
except for the larger $x$ region in the case of L3 data. 
In the present analysis, we have treated the active flavors as massless
quarks, and ignored the mass effects of the heavy flavors, which should
remain as a future subject. We should also investigate the power corrections
due to the higher-twist effects as well as possible resummation of large
logs as $x$ approaches $x_{\rm max}$.
%We expect the future experiments would provides us with more accurate
%data for the double-tag two-photon processes in $e^+ e^-$ collisions.

\acknowledgments

We would like to thank the organizers of the RADCOR 2007 for the
hospitality at such a well-organized and stimulating symposium.

%%%%%%%%%%%%%%%%%%%%%%%%%%%%%%%%%%%
\begin{figure*}
\centering
  \begin{tabular}{c@{\hspace*{15mm}}c}
    \includegraphics[scale=0.5]{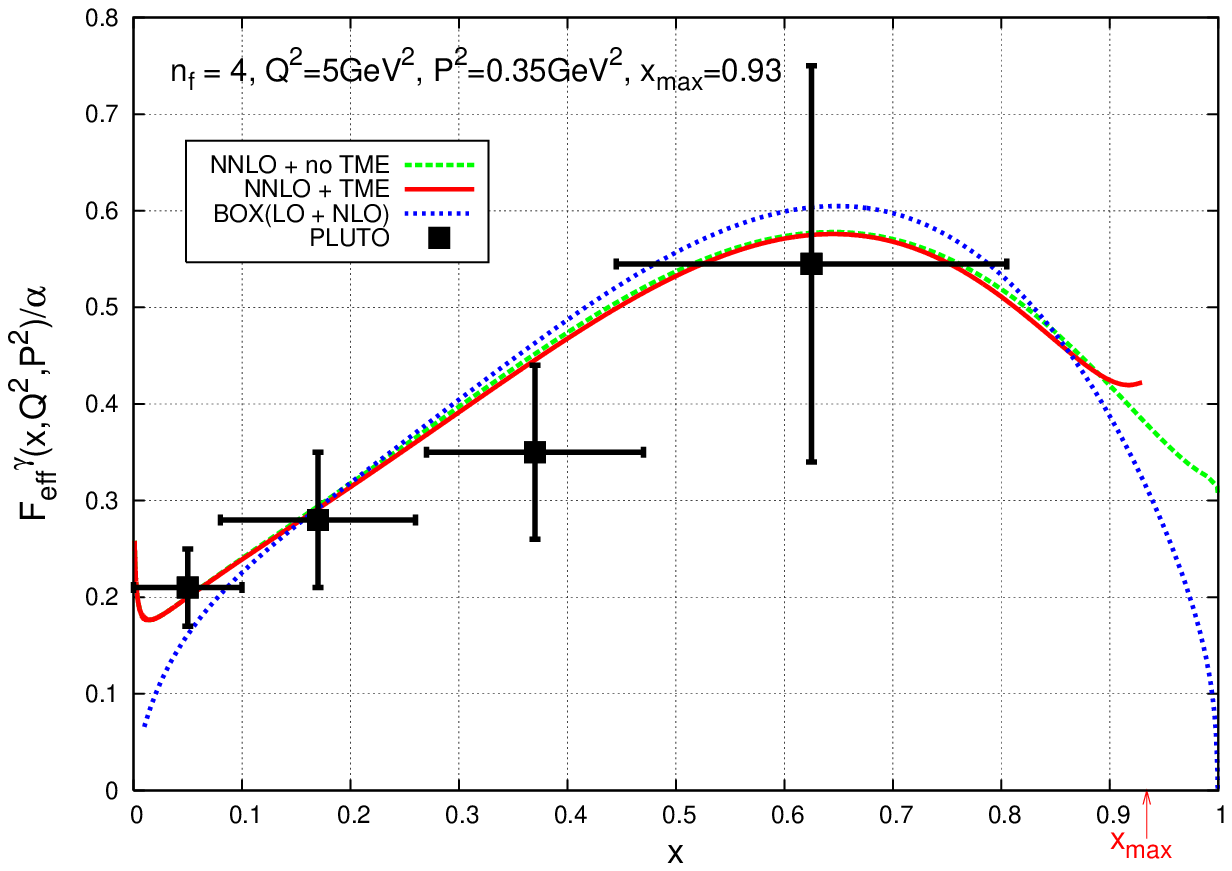}
    &
   \includegraphics[scale=0.5]{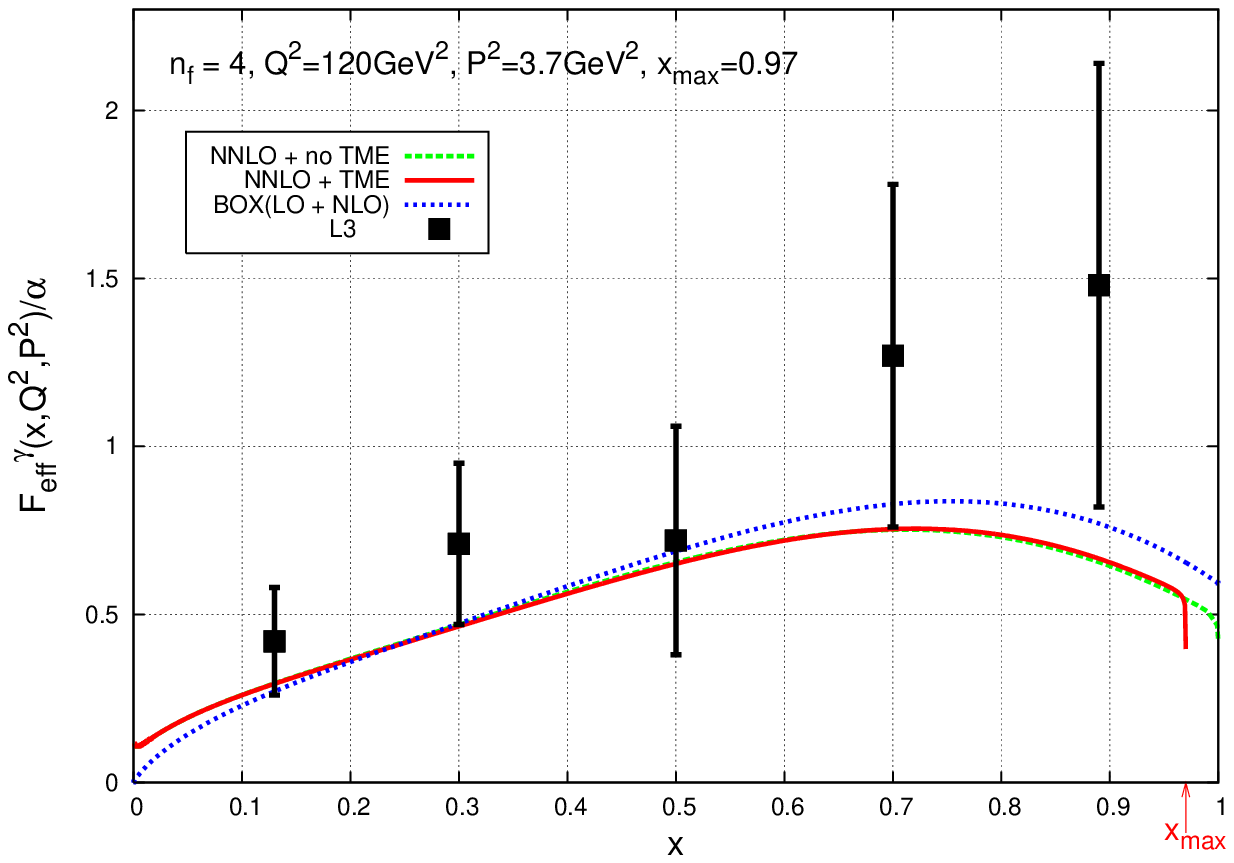}\\
(a) & (b)
  \end{tabular}
\caption{(a) $F_{\rm eff}^\gamma(x,Q^2,P^2)$ and data from PLUTO \cite{PLUTO}
and (b) $F_{\rm eff}^\gamma(x,Q^2,P^2)$ and data from L3 \cite{L3}.
}
\label{fig3}
\end{figure*}
%%%%%%%%%%%%%%%%%%%%%%%%%%%%%%%%%%%

\end{document}